\documentstyle[twocolumn,aps]{revtex}
\begin{document}
\baselineskip=4.0mm
\def\a{\alpha}
\noindent
{\bf
Comment on "Hole-Burning Experiments within Glassy Models with Infinite Range
Interactions"\\
}

Using a model devoid of an explicit spatial structure, in a recent
Letter\cite{1} Cugliandolo and Iguain (CI) claim to be able to reproduce
several features of nonresonant spectral hole burning (NHB) thus suggesting
that NHB may not be suited to map out dynamic heterogeneity.
Here we will show that the results presented by CI are not appropriate to
support such a claim.

CI consider the $p=3$ spherical spin-glass which is closely related to
Leutheusser's model of the structural glass transition\cite{2}. They report
pump frequency dependent distortions of the integrated response which they
interpret as evidence for frequency selectivity, one of the main features of
NHB. The model exhibits on-site (or rotational) dynamics only, thus precluding
a straightforward application to translational processes. In the experimental
studies addressing the rotational dynamics of glass-formers\cite{3} frequency
selective NHB signals were properly presented as evidence for {\it dynamic}
heterogeneities of the $\a$-relaxation\cite{4}.
We have emphasized previously that\cite{3} "it is not fully established that
{\it spatial} heterogeneity is at the origin" of these observations.
In this respect CI do not cite our work correctly.

Unfortunately CI have not followed the standard protocol of NHB which requires
to establish metastable equilibrium prior to the pump process\cite{3}.
Merely, CI perform infinitely fast quenches from $T=\infty$ to temperatures
$T=0.8$ and $T=0.59$. At $T=0.8$ the waiting time is chosen as $t_w=0$. Thus,
at least at short pump durations, $t_1$, the out-of-equilibrium dynamics
present in this model interfere with possible NHB effects. This is because
for $T>T_c(p=3)\sim 0.61$\cite{1} the fluctuation dissipation theorem is
violated on the time scale $t_\a$ of the $\a$-relaxation\cite{5}.
$t_\a$ tends to diverge upon approaching $T_c$. Therefore, the data
reported for $T=0.59$ again represent out-of-equilibrium dynamics and not
the $\a$-relaxation as required for a direct comparison with the experimental
results\cite{3}. Interestingly, upon increasing $t_w$ the apparent frequency
selectivity seen for the lower temperature in Fig. 7 of Ref. 1 diminishes.
Thus a scenario is approached for which $t_1$ dependent NHB signals exhibit
the same shape and only differ in their amplitudes.

An absence of frequency selectivity is the hallmark of homogeneous
dynamics\cite{3,6}.
Observations of homogeneous relaxations for the $p=3$ model would not be
surprising, since (for $T>T_c$) the $\a$-relaxation of this model is known to
proceed in an exponential fashion\cite{7}.
At a given $T$ this implies the existence of a unique $\a$-relaxation time
which in turn precludes observation of heterogeneity in the $\a$-response of
this model. However, the interpretation of the effects reported at $T=0.8$ is
not only hampered by the presence of out-of-equilibrium dynamics, which could
be removed by a suitable choice of $t_w$, but additionally by the fact that at
$T=0.8$ $\a$-process and high-frequency relaxations take place on about the
same time scale. If out-of-equilibrium effects can be eliminated, it remains
an interesting question whether the short-time as well as the long-time
behaviors of specific models imply dynamical or even spatial heterogeneities.

In summary the study of CI reveals that out-of-equilibrium dynamics can produce
phenomena which, without proper caution, can be confused with those
characteristic of NHB. Hence, the results so far presented by CI do not touch
upon the conclusion that the supercooled liquids studied previously do exhibit
unequivocal evidence for dynamic heterogeneity\cite{3,4}. We should
emphasize that we do not claim that frequency selectivity is
generally ruled out in models involving quenched disorder, irrespective of
whether or not they exhibit a spatial structure.

\vspace{0.4cm}
\noindent
Gregor Diezemann and Roland B\"ohmer\\
Institut f\"ur Physikalische Chemie,\\
Universit\"at Mainz, 55099 Mainz, FRG

\vspace{0.4cm}
\noindent
Received \hspace{1.5cm} December 2000

\vspace{0.4cm}

\end{document}